\documentstyle[floats,prl,aps]{revtex}
\input epsf
\begin{document}

\newcommand{\vertsp}{\vphantom{\displaystyle{\dot a \over a}}}
\newcommand{\se}{{(0)}}
\newcommand{\ve}{{(1)}}
\newcommand{\te}{{(2)}}
\newcommand{\nnu}{\nu}
\newcommand{\Spy}[3]{\, {}_{#1}^{\vphantom{#3}} Y_{#2}^{#3}}
\newcommand{\Gm}[3]{\, {}_{#1}^{\vphantom{#3}} G_{#2}^{#3}}
\newcommand{\Spin}[4]{\, {}_{#2}^{\vphantom{#4}} {#1}_{#3}^{#4}}
\newcommand{\scpot}{{\cal V}}

 \renewcommand{\ell}{l}  
 \renewcommand{\topfraction}{1.0}
\renewcommand{\bottomfraction}{1.0}
\renewcommand{\textfraction}{0.00}
\renewcommand{\dbltopfraction}{1.0}

\title{A COMPLETE TREATMENT OF \\
CMB ANISOTROPIES IN A FRW UNIVERSE}

\author{Wayne Hu${}^1$\footnote{Alfred P. Sloan Fellow}, 
 	Uro\v s Seljak${}^2$,
	Martin White${}^3$, \&
	Matias Zaldarriaga${}^4$}
 
\address{${}^1$ Institute for Advanced Study, School of Natural Sciences,
	Princeton, NJ 08540  \\
	${}^2$ Harvard Smithsonian Center For Astrophysics, Cambridge, 
	MA 02138 \\
        ${}^3$Departments of Physics and Astronomy, University of Illinois
	at Urbana-Champaign, Urbana, IL 61801\\ 
	${}^4$ Department of Physics, MIT, Cambridge, MA 02139 } 

\maketitle

\begin{abstract}
We generalize the total angular momentum method for computing Cosmic 
Microwave Background anisotropies to  Friedman-Robertson-Walker (FRW) 
spaces with arbitrary geometries.
This unifies the treatment of temperature and polarization anisotropies
generated by scalar, vector and tensor perturbations of the fluid, seed,
or a scalar field, in a universe with constant comoving curvature.
The resulting formalism generalizes and simplifies the calculation of
anisotropies and, in its integral form, allows for a fast calculation of
model predictions in linear theory for any FRW metric.
\end{abstract}


\section{Introduction}

The study of the Cosmic Microwave Background (CMB) radiation holds the key
to understanding the seeds of the structure we see around us in the universe,
and could potentially enable precision measures for most of the important
cosmological parameters.
For this reason, as well as because of its intrinsic interest, one would
like a physically transparent framework for the study of CMB anisotropies
which is as general, powerful, and flexible as possible.

Theoretically, the calculation of CMB anisotropies is ``clean'', involving as
it does only linear perturbation theory.
However the calculations can become quite complex once one allows for the
possibility of non-flat universes, non-scalar perturbations to the metric,
and polarization as well as temperature anisotropies.
Recently Hu \& White \cite{TAMM} presented a formalism for calculating CMB
anisotropies which treats all types of perturbations, 
temperature and polarization anisotropies, 
and hierarchy and integral solutions on an equal footing.  
The formalism, named the total angular momentum method,  greatly simplifies
the physical interpretation of the equations and the form of their solutions
(see e.g.~\cite{Polar}).
However it was presented in detail only for the case of flat spatial
hypersurfaces.  Here we generalize the treatment for the curved spaces
of open and closed Friedman-Robertson-Walker (FRW) universes.

Aspects of this method in open (hyperbolic, negatively curved) geometries have 
been introduced
in Hu \& White \cite{OpenTen} and Zaldarriaga, Seljak \& Bertschinger
\cite{ZalSelBer} for the cases of tensor temperature and scalar polarization
respectively.  
The latter work also addressed methods for efficient implementation
through the line of sight integration technique \cite{LOS}.
In this paper, we complete the total angular momentum method for 
arbitrary perturbation type and
FRW metric, paying particular attention to the case of open 
universes because of its strong observational motivation. 
As an example we use this formalism to compute the temperature and polarization
angular power spectra of both scalar and tensor modes in critical density and
open inflationary models. 
We incorporated the formalism into the CMBFAST code of Seljak \& Zaldarriaga
\cite{LOS}, which has been made publically available.

The outline of the paper is as follows:
we begin by establishing our notation for fluctuations about a FRW background
cosmology in \S\ref{sec:metric}.
We then present the Boltzmann equation in our formalism in
\S\ref{sec:boltzmann}, which contains the main results.
We give some examples and discuss applications in \S\ref{sec:discussion}.
Some of the more technical parts of the derivations (the Einstein, radial
and hierarchy equations) are presented in a series of three Appendices.

\section{Metric and Stress-Energy Perturbations} \label{sec:metric}

In this section, we discuss the representation of the perturbations for
the cosmological fluids and the geometry of space-time.
We start by defining the basis in which we shall expand such perturbations
and their representation under various gauge choices.

We assume that the background is described by an FRW metric 
$g_{\mu\nu} = a^2 \gamma_{\mu\nu}$ 
with scale factor $a(t)$ and constant comoving curvature
$K = - H_0^2(1-\Omega_{\rm tot})$ in the spatial metric $\gamma_{ij}$.
Here greek indices run from $0$ to $3$ while latin indices run over the
spatial part of the metric: $i,j=1,2,3$.
It is often convenient to represent the metric in spherical coordinates where 
\begin{equation}
\gamma_{ij} dx^i dx^j = |K|^{-1} \left[ d\chi^2 + \sin_K^2\chi 
	( d \theta^2 + \sin^2\theta\, d\phi^2 ) \right]\,, 
\end{equation}
with
\begin{equation}
\sin_K(\chi) = \cases { \sinh(\chi)\,, & $K<0\,,$ \cr
			\sin(\chi)\,, & $K>0\,,$ \cr}
\end{equation} 
where the flat-limit expressions are regained as $K \rightarrow 0$ from
above or below.  The component corresponding to conformal time 
\begin{equation}
x^0 \equiv \eta = \int {dt\over a(t)}
\end{equation}
is $\gamma_{00}=-1$.

Small perturbations $h_{\mu\nu}$ around this FRW metric
\begin{equation}
g_{\mu\nu} = a^2(\gamma_{\mu\nu} + h_{\mu\nu})\,,
\end{equation}
can be decomposed into
scalar ($m=0$, compressional),
vector ($m=\pm 1$, vortical) and tensor ($m=\pm 2$, gravitational wave)
components from their transformation properties under spatial rotations
\cite{AbbSch,TAMM}.

\subsection{Eigenmodes}
\label{sec:eigenmodes}

In linear theory, each eigenmode of the Laplacian for the perturbation
evolves independently, and so it is useful to decompose the perturbations
via the eigentensor ${\bf Q}^{(m)}$, where
\begin{equation}
\nabla^2 {\bf Q}^{(m)} \equiv \gamma^{ij} {\bf Q}_{|ij}^{(m)} = 
-k^2 {\bf Q}^{(m)} ,
\end{equation}
with ``$|$'' representing covariant differentiation with respect to
the three metric $\gamma_{ij}$.
Note that the eigentensor ${\bf Q}^{(m)}$ has $|m|$ indices (suppressed
in the above).  Vector and tensor modes also satisfy the auxiliary conditions
\begin{eqnarray}
Q^{(\pm 1)}_i{}^{|i} &=& 0\, , \nonumber\\
\gamma^{ij} Q^{(\pm 2)}_{ij} &=& Q^{(\pm 2)}_{ij}{}^{|i} = 0 \,,
\end{eqnarray}
which represent the divergenceless and transverse-traceless conditions
respectively, as appropriate for vorticity and gravity waves.
In flat space, these modes are particularly simple and may be expressed as
\begin{equation}
Q_{i_1 \ldots i_{m}}^{(\pm m)} \propto 
	(\hat{e}_1 \pm i \hat{e}_2)_{i_1} \ldots 
	(\hat{e}_1 \pm i \hat{e}_2)_{i_m}
	\exp(i \vec{k} \cdot \vec{x})\,, \qquad (K=0, m\ge 0)\,,
\end{equation} 
where the presence of $\hat{e}_i$, which forms  
a local orthonormal basis with $\hat{e}_3=\hat{k}$, ensures the 
divergenceless and transverse-traceless conditions.

It is also useful to construct (auxiliary) vector and tensor objects out
of the fundamental scalar and vector modes through covariant differentiation
\begin{equation}
Q_i^{(0)} = -k^{-1} Q_{|i}^{(0)}\,, \qquad Q_{ij}^{(0)} 
	= k^{-2} Q_{|ij}^{(0)} + {1 \over 3} \gamma_{ij} Q^{(0)} \,,
\end{equation}
\begin{equation}
Q^{(\pm 1)}_{ij} = -(2k)^{-1}( Q^{(\pm 1)}_{i|j} + Q^{(\pm 1)}_{j|i} ).
\end{equation}

The completeness properties of these eigenmodes are 
discussed in detail in
\cite{AbbSch}, where it is shown that in terms of the generalized wavenumber
\begin{equation}
q = \sqrt{k^2+(|m|+1)K} \, , \qquad  \nnu = q/|K|\,,
\end{equation}
the spectrum is complete for 
\begin{equation}
\begin{array}{rll}
\nnu  &\ge  0, \qquad & K<0\,, \\
      & =   3,4,5\ldots, \qquad &K>0\,.
\end{array}
\end{equation}
A deceptive aspect of this labelling is that for an open universe
the characteristic scale of
the structure in a mode is $2\pi/k$ and {\it not\/} $2\pi/q$, so all
functions have structure only out to the curvature scale even as
$q \rightarrow 0$.  
We often go between the variable sets $(k,\eta)$, $(q,\eta)$ and
$(\nnu,\chi)$ for convenience. 

\subsection{Perturbation Representation}

A general metric perturbation can be broken up into the normal modes of
scalar ($m=0$), vector ($m=\pm 1$) and tensor ($m=\pm 2)$ types,
\begin{eqnarray}
h_{00} &=& - \sum_m 2 A^{(m)} Q^{(m)} \,, \nonumber \\
h_{0i} &=& - \sum_m B^{(m)} Q_i^{(m)} \,, \nonumber \\
h_{ij} &=& \sum_m 2 H_L^{(m)} Q^{(m)} \gamma_{ij}+2 H_T^{(m)} Q_{ij}^{(m)} \,. 
\end{eqnarray}
Note that scalar quantities cannot be formed from vector and tensor modes
so that $A^{(m)}=0$ and $H_L^{(m)}=0$ for $m\ne 0$; likewise vector quantities
cannot be formed from tensor modes so that $B^{(m)}=0$ for $|m| = 2$.

There remains gauge freedom associated with the coordinate
choice for the metric perturbations (see Appendix \ref{sec:gauge}).
It is typically employed to eliminate two out of four of these
quantities for scalar perturbations and one of the two for vector
perturbations.  The metric is thus specified by four quantities.  
Two popular choices are the {\it synchronous\/} gauge, where
\begin{eqnarray}
H_L^{(0)} = h_L, &\qquad& H_T^{(0)} = h_T , \nonumber\\
H_T^{(1)} = h_V, &\qquad& H_T^{(2)} = H,
\end{eqnarray}
and the generalized (or conformal) {\it Newtonian\/} gauge, where
\begin{eqnarray}
A^{(0)} = \Psi\,, &\qquad& B^{(1)}=V \,, \nonumber \\
H_L^{(0)} = \Phi\,, &\qquad& H_T^{(2)}=H \, .
\end{eqnarray}
Here and below, when only the $m\ge 0$ expressions are displayed, the
$m < 0$ expressions should be taken to be identical unless otherwise
specified.

The stress energy tensor can likewise be broken up into scalar, vector,
and tensor contributions.  Furthermore one can separate fluid ($f$)
contributions and seed ($s$) contributions.
The latter is distinguished by the fact that the net effect can be viewed
as a perturbation to the background.
Specifically $T_{\mu\nu} = \bar T_{\mu\nu} + \delta T_{\mu\nu}$ where
$\bar T^0_{\hphantom{0}0} = -\rho_f$,
$\bar T^0_{\hphantom{0}i} = \bar T_0^{\hphantom{i}i} =0$
and $\bar T^i_{\hphantom{i}j} = p_f \delta^i_{\hphantom{i}j}$
is given by the fluid alone.
The fluctuations can be decomposed into the normal modes of
\S\ref{sec:eigenmodes} as
\begin{equation}
\begin{array}{lcl}
\delta T^0_{\hphantom{0}0} &=&
        - \sum_m  [\rho_f \delta_f^{(m)} + \rho_s] \, Q^{(m)}, \vertsp\\
\delta T^0_{\hphantom{0}i} &=& \sum_m  [(\rho_f + p_f) (v_f^{(m)}-B^{(m)})
        + v_s^{(m)}]
        \, Q_i^{(m)}  ,
        \vertsp\\
\delta T_0^{\hphantom{i}i} &=&
	 -\sum_m [(\rho_f + p_f)v_f^{(m)} + v_s^{(m)}] \, Q^{(m)}{}^i,
        \vertsp\\
\delta T^i_{\hphantom{i}j} &=& \sum_m [\delta p_f^{(m)} + p_s]
        \delta^i_{\hphantom{i}j} Q^{(m)} + [ p_f\pi_f^{(m)} + p_s]
        Q^{(m)}{}^i_{\hphantom{i}j} \vertsp \, .
\end{array}
\label{eqn:stress}
\end{equation}
Since $\delta^{(m)}_f=\delta p^{(m)}_f= 0$ for $m \ne 0$, we hereafter
drop the superscript from these quantities.  

A minimally coupled scalar field $\varphi$ with Lagrangian
\begin{equation}
{\cal L} = -{1 \over 2} \sqrt{-g} \left[ g^{\mu \nu} \partial_\mu \varphi 
	\partial_\nu \varphi + 2V(\varphi) \right]
\end{equation}
can be treated in the same way with the associations
\begin{equation}
\rho_\phi = p_\phi + 2\scpot =
   {1 \over 2} a^{-2} \dot \phi^2 + \scpot \,,
\label{eqn:scalarfieldbac}
\end{equation}
for the background density and pressure.  The fluctuations $\varphi = \phi+ \delta\phi$
are related to the fluid quantities as \cite{KodSas}
\begin{eqnarray}
\delta \rho_\phi =\delta p_\phi + 2\scpot_{,\phi}\delta\phi &=& a^{-2}(\dot\phi 
\dot{\delta\phi}-A^{(0)} \dot\phi^2)+\scpot_{,\phi} \delta\phi \nonumber\,,\\
(\rho_\phi+p_\phi) (v^{(0)}_\phi- B^{(0)}) &=& a^{-2} k \dot\phi \delta\phi 
	\nonumber\,,\\
p_\phi \pi_\phi^{(0)} & = & 0\,.
\label{eqn:scalarfieldpert}
\end{eqnarray}

The evolution of the matter and metric perturbations follows from the
Einstein equations $G_{\mu\nu} = 8\pi G T_{\mu\nu}$ and encorporates
the continuity and Euler equations through the implied energy-momentum
conservation $T^{\mu\nu}{}_{;\nu} = 0$.
We give these relations explicitly for the scalar, vector and tensor
perturbations in both Newtonian and synchronous gauge in
Appendix \ref{sec:einstein} (see also \cite{HSW}).

These equations hold equally well for relativistic matter such as the
CMB photons and the neutrinos.  However in that case they do not represent
a closed system of equations (the equation of motion of the anisotropic
stress perturbations $\pi_f^{(m)}$ is unspecified) and do not account for
the higher moments of the distribution or for momentum exchange between
different particle species.
To include these effects, we require the Boltzmann equation
which describes the evolution of the full distribution function under
collisional processes. 

\section{Boltzmann Equation}
\label{sec:boltzmann}

The Boltzmann equation describes the evolution in time $(\eta)$ of the 
spatial ($\vec{x}$) and angular ($\hat{n}$) distribution of the radiation
under gravity and scattering processes.  In the notation of \cite{TAMM}, it
can be written implicitly as
\begin{equation}
{d \over d\eta} \vec{T}(\eta,\vec{x},\hat{n}) \equiv 
{\partial \over \partial\eta} \vec{T} + n^i \vec{T}_{|i} 
= \vec{C}[\vec{T}] + \vec{G}[h_{\mu\nu}] \, ,
\label{eqn:boltzmannimplicit}
\end{equation}
where $\vec{T} = (\Theta, Q+iU, Q-iU)$ encapsulates the perturbation 
to the temperature $\Theta=\Delta T/T$ and the polarization
(Stokes $Q$ and $U$ parameters) in units of the temperature fluctuation.
The term $\vec{C}$ accounts for collisions, here  Compton scattering of the
photons with the electrons, 
while the term $\vec{G}$ accounts for gravitational redshifts.

\subsection{Metric and Scattering Sources}

The gravitational term $\vec{G}$
is easily evaluated from the Euler-Lagrange equations
for the motion of a massless particle in the background given by $g_{\mu\nu}$
\cite{AbbSch,SacWol,WSS}:
\begin{equation}
\vec{G}[h_{\mu\nu}] = \left({1 \over 2}
         {n}^i {n}^j \dot h_{ij}
+ {n}^i \dot h_{0i}
        + {1 \over 2} n^i h_{00|i} \,,  0 \, , 0 \right) .
\label{eqn:gravred}
\end{equation}
Note that gravitational redshifts affect different polarization states alike. 
As should be expected, the modification from the flat space case involves the
replacement of ordinary spatial derivatives with covariant ones. 

The Compton scattering term $\vec{C}$ was derived in \cite{TAMM,ZalSelBer}
in the total angular momentum language.
Though the basic result has long been known \cite{Cha,BonEfs}, this
representation has the virtue of explicitly showing that complications due
to the angular and polarization dependence of Compton scattering come simply
through the quadrupole moments of the distribution. Here
\begin{eqnarray}
\vec{C}[{\vec {T}}] &=& -\dot\tau 
	\left[ \vec{T}(\hat{n})
 - \left( \int {d\hat{n}'\over 4\pi}
        \Theta' + \hat{n} \cdot \vec{v}_B \, , 0 , \, 0 \right) 
	\right] 
	+
        {\dot\tau \over 10} \int d\hat{n}'
        \sum_{m=-2}^2 {\bf P}^{(m)}(\hat{n},\hat{n}') \vec{T}(\hat{n}')\, ,
\label{eqn:fullcollision}
\end{eqnarray}
where the differential cross section for Compton scattering is
$\dot\tau = n_e \sigma_T a$ where $n_e$ is the free electron number density
and $\sigma_T$ is the Thomson cross section.
The bracketed term in the collision integral describes the isotropization of
the photons in the rest frame of the electrons. 
The last term accounts for the angular and polarization dependence of the
scattering with
\begin{eqnarray}
{\bf P}^{(m)} =
\left(
\begin{array}{ccc}
Y_2^{m}{}'\, Y_2^m   \quad &
          - \sqrt{3 \over 2} \Spy{2}{2}{m}{}'\, Y_2^m
        \quad & - \sqrt{3 \over 2}
          \Spy{-2}{2}{m}{}'\, Y_2^m
        \vertsp\\
- \sqrt{6} Y_2^{m}{}' \Spy{2}{2}{m} \quad &
3 \Spy{2}{2}{m}{}'\Spy{2}{2}{m} \quad &
          3 \Spy{-2}{2}{m}{}'\Spy{2}{2}{m}{} \quad
                \vertsp\\
- \sqrt{6} Y_2^{m}{}' \Spy{-2}{2}{m}{} \quad &
          3 \Spy{2}{2}{m}{}' \Spy{-2}{2}{m} \quad &
          3 \Spy{-2}{2}{m}{}' \Spy{-2}{2}{m} ,
                \vertsp\\
\end{array}
\right),
\label{eqn:scatmatrix}
\end{eqnarray}
where $Y_\ell^m {}'\equiv Y_\ell^{m*} (\hat{n}')$ and
$\Spy{s}{\ell}{m}{}'\equiv\Spy{s}{\ell}{m*}(\hat{n}')$ and the unprimed
harmonics have argument $\hat{n}$.
Here $\Spin{Y}{s}{\ell}{m}$ are the spin-weighted spherical harmonics
\cite{Spin,SelZal,KamKosSte,TAMM}.

\subsection{Normal Modes} 
\label{sec:normal}

The temperature and polarization distributions are functions of the position
$\vec{x}$ and the direction of propagation of the photons $\vec{n}$.  
They can be expanded in modes which account for both the local angular and
spatial variations: $\Spin{G}{s}{\ell}{m}(\vec{x},\hat{n})$, i.e.
\begin{equation}
\begin{array}{rcl}
\Theta(\eta,\vec{x},\hat{n}) &=& \displaystyle{
\int {d^3 q \over (2\pi)^3} }
        \sum_{\ell}
\sum_{m=-2}^2 \Theta_\ell^{(m)} \Gm{0}{\ell}{m} \, , \\
 (Q \pm i U)(\eta,\vec{x},\hat{n}) &=& \displaystyle{\int {d^3q
\over (2\pi)^3}}
        \sum_{\ell} \sum_{m=-2}^2
        (E_\ell^{(m)} \pm i B_\ell^{(m)}) \, \Gm{\pm 2}{\ell}{m} \,, 
\end{array}
\label{eqn:decomposition}
\end{equation}
with spin $s=0$ describing the temperature fluctuation and $s=\pm 2$
describing the polarization tensor.
$E_\ell$ and $B_\ell$ are the angular moments of the electric and magnetic
polarization components.
It is apparent that the effects of the local scattering process $\vec{C}$
is most simply evaluated in a representation where the separation of the
local angular and spatial distribution is explicit \cite{TAMM}, with the
former being an expansion in $\Spy{s}{\ell}{m}$.
The subtlety lies in relating the local basis at two {\it different\/}
coordinate points, say the last scattering event and the observer.

In flat space, the representation is straightforward since the parallel
transport of the angular basis in space is trivial.  The result is a product
of spin-weighted harmonics for the local angular dependence and plane waves
for the spatial dependence:
\begin{equation}
\Gm{s}{\ell}{m}(\vec{x},\hat{n})  =
        (-i)^\ell \sqrt{ {4\pi \over 2\ell+1}}
        [\Spy{s}{\ell}{m}(\hat{n})] \exp(i\vec{k} \cdot \vec{x})\,,
	\qquad (K=0)\,.
\label{eqn:flatG}
\end{equation}
Here we seek a similar construction in an curved geometry.  
We will see that this construction greatly simplifies the scalar harmonic
treatment of \cite{Wil,WhiSco,ZalSelBer} and extends it to vector and
tensor temperature \cite{OpenTen} modes as well as all polarization modes.

To generalize these modes to the curved geometry, we wish to replace the
plane wave with some spatially dependent phase factor
$\exp[i \delta(\vec{x},\vec{k})]$ related to the eigenfunctions
${\bf Q}^{(m)}$  of \S\ref{sec:eigenmodes} while keeping the same local
angular dependence (see Eq.~\ref{eqn:generalizedGA}). 
By virtue of this requirement, the Compton scattering terms, which
involve only the local angular dependence, retain the same form as in
flat space. 
In Appendix \ref{sec:derivation}, we derive $\Spin{G}{s}{\ell}{m}$ by
recursion from covariant contractions of the fundamental basis ${\bf Q}^{(m)}$.
The result is a recursive definition of the basis
\begin{equation}
n^i (\Spin{G}{s}{\ell}{m})_{|i} 
        = {q \over 2\ell +1} \left[
          {\Spin{\kappa}{s}{\ell}{m}}  (\Spin{G}{s}{\ell-1}{m})
        - {\Spin{\kappa}{s}{\ell+1}{m}} (\Spin{G}{s}{\ell+1}{m})
        \right]
        - i{q m s \over \ell(\ell+1)}\ \Spin{G}{s}{\ell}{m} \, ,
\label{eqn:recursion}
\end{equation}
constructed from the lowest $\ell$-mode of Eq.~(\ref{eqn:Gl0})
with the coupling coefficient
\begin{equation}
\Spin{\kappa}{s}{\ell}{m} = \sqrt{ \left[
{(\ell^2-m^2)(\ell^2-s^2)\over\ell^2}\right]
\left[1 - {\ell^2\over q^2} K \right]}.
\label{eqn:coupling}
\end{equation}
The structure of this relation is readily apparent.  The recursion relation
expresses the addition of angular momentum and is the defining equation in
the total angular momentum method.
It says the ``total'' local angular dependence at (say) the origin is the
sum of the local angular dependence at distant points (``spin'' angular
momentum) plus the angular variations induced by the spatial dependence of
the mode (``orbital'' angular momentum).

The recursion relation represents the addition of angular momentum for the
case of an infinitesimal spatial separation. 
Here the leading order spatial variation is the gradient
[$n^i (\Spin{G}{s}{\ell}{m})_{|i}$] 
term which has an angular structure of a dipole $Y_1^0$.
The first term on the rhs of equation~(\ref{eqn:coupling}) arises from the
Clebsch-Gordan relation that couples the orbital $Y_1^0$ with the intrinsic
$\Spin{Y}{s}{\ell}{m}$ to form $\ell\pm 1$ states,
\begin{eqnarray}
\sqrt{4 \pi \over 3} Y_1^0 (\Spin{Y}{s}{\ell}{m})
    &=&  { \Spin{c}{s}{\ell}{m} \over \sqrt{(2\ell+1)(2\ell-1)}}
         \left(\Spin{Y}{s}{\ell-1}{m}\right)
+ {\Spin{c}{s}{\ell+1}{m} \over \sqrt{(2\ell+1)(2\ell+3)}}
        \left(\Spin{Y}{s}{\ell+1}{m}\right) 
	- {m s \over \ell (\ell +1)} \left({}_s Y_{\ell}^m\right) \,,
\label{eqn:streamingcg}
\end{eqnarray}
where the coupling coefficient is $\Spin{c}{s}{\ell}{m} = 
\sqrt{(\ell^2-m^2)(\ell^2-s^2)/\ell^2}$.

The second term on the rhs of the coupling equation (\ref{eqn:coupling}) 
accounts for geodesic deviation factors in the conversion of spatial
structure into orbital angular momentum.
Consider first a closed universe with radius of curvature ${\cal R}=K^{-1/2}$.
Suppressing one spatial coordinate, we can analyze the problem as geometry on
the 2-sphere with the observer situated at the pole.
Light travels on radial geodesics or great circles of fixed longitude.
A physical scale $\lambda$ at fixed latitude (given by the polar angle $\chi$)
subtends an angle $\alpha = \lambda/{\cal R}\sin\chi$.
In the small angle approximation, a Euclidean analysis would infer a distance
related by
\begin{equation}
{\cal D}(d) = {\cal R} \sin \chi =
K^{-1/2} \sin \chi  \, ,    \qquad (K>0),
\end{equation}
called here the {\it angular diameter distance}.
For negatively curved or open universes, a similar analysis implies
\begin{equation}
{\cal D}(d) = |K|^{-1/2} \sinh \chi\, ,
\qquad (K<0).
\label{eqn:angularsize}
\end{equation}
Thus the angular scale corresponding to an eigenmode of
wavelength $\lambda$ is
\begin{equation}
\theta = {\lambda \over {\cal R} \sinh{\chi}} \, 
       \approx {1 \over \nnu \sinh{\chi}} \, .
\end{equation}
For an infinitesimal change $\chi$, orbital angular momentum
of order $\ell$ is stimulated when 
\begin{eqnarray}
\chi &\approx& {1\over\nu\theta} [1+{\cal O}(\nu^2\theta^2)]\,, \nonumber\\
\eta &\approx&  {\ell\over q}[1+{\cal O}(\ell^2 K / q^2) ]\,,
\end{eqnarray}
which explains the factors of $\ell^2 K / q^2$ in the coupling term in a
curved geometry.  
We shall see in \S\ref{sec:integral} that these infinitesimal additions of
angular momentum and geodesic deviation may be encorporated into a single
step by finding the integral solutions to the coupling equation
(\ref{eqn:recursion}).

\subsection{Evolution Equations} 

It is now straightforward to rewrite the  Boltzmann equation
(\ref{eqn:boltzmannimplicit}) as the evolution equations for the amplitudes
of the normal modes of the temperature and polarization
$\vec{T}_\ell^{(m)} = (\Theta^{(m)}_\ell, E^{(m)}_\ell, B_\ell^{(m)})$.
The gravitational sources and scattering sources of these equations
follow from Eq. (\ref{eqn:gravred}) and (\ref{eqn:fullcollision}) by noting
that the spin harmonics are orthogonal,
\begin{equation}
\int d\Omega\ (\Spin{Y}{s}{\ell}{m})(\Spin{Y}{s}{\ell'}{m'}{}^*)
        = \delta_{\ell,\ell'} \delta_{m m'}.
\end{equation}
The term $n^i \vec{T}_{|i}$ is evaluated by use of the coupling relation
Eq.~(\ref{eqn:recursion}) for $n^i(\Spin{G}{s}{\ell}{m})_{|i}$.
It represents the fact that spatial gradients in the distribution become
orbital angular momentum as the radiation streams along its trajectory
$\vec{x}(\hat{n})$.
For example, a temperature variation on a distant surface surrounding the
observer appears as an anisotropy on the sky.
This process then simply reflects a projection relation that relates distant
sources to present day local anisotropies.

With these considerations, the temperature fluctuation evolves as
\begin{equation}
\dot\Theta_\ell^{(m)}
= q     \Bigg[ {\Spin{\kappa}{0}{\ell}{m} \over (2\ell-1)}
        \Theta_{\ell-1}^{(m)}
             -{\Spin{\kappa}{0}{\ell+1}{m}
                \over (2\ell+3)}
        \Theta_{\ell+1}^{(m)} \Bigg]
	- \dot\tau \Theta_\ell^{(m)} + S_\ell^{(m)},  
	\qquad (\ell \ge m),
\label{eqn:boltz}
\end{equation}
and the polarization as
\begin{eqnarray}
\dot E_\ell^{(m)} &=&
        q
	\Bigg[ {\Spin{\kappa}{2}{\ell}{m} \over (2\ell-1)}
E_{\ell-1}^{(m)} - {2m \over \ell (\ell + 1)} B_\ell^{(m)} 
- {\Spin{\kappa}{2}{\ell+1}{m} \over (2 \ell + 3)}
        E_{\ell + 1}^{(m)} \Bigg] 
	- \dot\tau [E_\ell^{(m)}+\sqrt{6}P^{(m)}\delta_{\ell,2}]\,,\nonumber\\
\dot B_\ell^{(m)} &=&
        q \Bigg[ {\Spin{\kappa}{2}{\ell}{m} \over (2\ell-1)}
B_{\ell-1}^{(m)} + {2m \over \ell (\ell + 1)} E_\ell^{(m)} 
- {\Spin{\kappa}{2}{\ell+1}{m} \over (2 \ell + 3)}
        B_{\ell + 1}^{(m)} \Bigg] -\dot\tau B_\ell^{(m)}.
\label{eqn:boltzpol}
\end{eqnarray}

The temperature fluctuation sources in Newtonian gauge are
\begin{equation}
\begin{array}{lll}
S_0^{(0)} = \dot\tau \Theta_0^{(0)} -  \dot\Phi \, ,      \qquad &
S_1^{(0)} = \dot\tau v_B^{(0)} + k\Psi \, ,               \qquad &
S_2^{(0)} = \dot\tau P^{(0)} \, , \vertsp\\
                                                \qquad &
S_1^{(1)} = \dot\tau v_B^{(1)} + \dot V \, ,              \qquad &
S_2^{(1)} = \dot\tau P^{(1)} \, , \vertsp\\
                                                \qquad &
                                                \qquad &
S_2^{(2)} = \dot\tau P^{(2)} - \dot H      \vertsp \, ,
\end{array}
\end{equation}
and in synchronous gauge, 
\begin{equation}
\begin{array}{lll}
S_0^{(0)} = \dot\tau \Theta_0^{(0)} -  \dot h_L \, ,      \qquad &
S_1^{(0)} = \dot\tau v_B^{(0)} \, ,               \qquad &
S_2^{(0)} = \dot\tau P^{(0)}  -{2 \over 3}\sqrt{1-3K/k^2}\ \dot h_T
	\, , \vertsp\\
                                                \qquad &
S_1^{(1)} = \dot\tau v_B^{(1)} \, ,              \qquad &
S_2^{(1)} = \dot\tau P^{(1)} -{\sqrt{3} \over 3}\sqrt{1-2K/k^2}\ \dot h_V
	\, , \vertsp\\
                                                \qquad &
                                                \qquad &
S_2^{(2)} = \dot\tau P^{(2)} - \dot H      \vertsp \, ,
\end{array}
\label{eqn:tempsources}
\end{equation}
The $\ell=m=2$ source doesn't contain a curvature factor because we have
recursively defined the basis functions in terms of the lowest member,
which is $\ell=2$ in this case.  In the above
\begin{equation}
P^{(m)} = {1 \over 10} \left[ \Theta_2^{(m)}  - \sqrt{6} E_2^{(m)} \right]
\label{eqn:polsource}
\end{equation}
and note that the photon density and velocities are related to the $\ell=0,1$
moments as
\begin{equation}
\delta_\gamma = 4\Theta_0^\se$\,, \qquad $v_\gamma^{(m)}=
\Theta_1^{(m)} \, ;
\end{equation}
whereas the anisotropic stresses are given by
\begin{equation}
\pi_\gamma^{(m)} Q_{ij}^{(m)} = 
12 \int {d \Omega \over 4\pi}\  (n_i n_j-{1\over 3}\gamma_{ij})\Theta^{(m)},
\end{equation}
which relates them to the quadrupole moments ($\ell=2$) as
\begin{equation}
(1-3K/k^2)^{1/2} \pi_\gamma^\se = {12 \over 5}\Theta_2^\se, \qquad
(1-2K/k^2)^{1/2} \pi_\gamma^\ve = {8\sqrt{3} \over 5} \Theta_2^\ve, 
\qquad \pi_\gamma^\te = {8\over 5} \Theta_2^\te .
\end{equation}
The evolution of the metric and matter sources are given in
Appendices \ref{sec:scalar}---\ref{sec:tensor}.  

\subsection{Integral Solutions}
\label{sec:integral}

The Boltzmann equations have formal integral solutions that are simple to
write down.  The hierarchy equations for the temperature distribution
Eq.~(\ref{eqn:boltz}) merely express the projection of the various plane
wave temperature sources $S_\ell^{(m)} \Gm{0}{\ell}{m}$ on the sky today
(see Eq.~(\ref{eqn:tempsources})).
Likewise Eq.~(\ref{eqn:boltzpol}) expresses the projection of
$-\sqrt{6} P^{(m)} \dot\tau e^{-\tau} \Spin{G}{\pm 2}{\ell}{m}$. 

The projection is obtained by extracting the total angular dependence of
the mode from its decomposition in spherical coordinates: i.e.~into radial
functions times spin harmonics $\Spin{Y}{s}{\ell}{m}$.
We discuss their explicit construction in Appendix \ref{sec:radial}.
The full solution immediately follows by integrating the projected source
over the radial coordinate,
\begin{eqnarray}
{\Theta_\ell^{(m)}(\eta_0,q) \over 2\ell + 1}\, & = & 
\int_0^{\eta_0} d\eta \ e^{-\tau} \, \sum_{j} S_{j}^{(m)} \, 
  \phi_{\ell}^{(jm)} ,\nonumber \\
{E^{(m)}_\ell(\eta_0,q) \over 2\ell+1} &=&  
\int_0^{\eta_0} d\eta \ \dot\tau e^{-\tau} (-\sqrt{6} P^{(m)}_{\vphantom{\ell}})
\,\epsilon_{\ell}^{(m)} ,\nonumber\\
{B^{(m)}_\ell(\eta_0,q) \over 2\ell+1} &=& 
\int_0^{\eta_0} d\eta \ \dot\tau e^{-\tau} (-\sqrt{6} P^{(m)}_{\vphantom{\ell}})
\,\beta_{\ell}^{(m)} ,
\label{eqn:los}
\end{eqnarray}
where the arguments of the radial functions
($\phi_\ell,\epsilon_\ell,\beta_\ell$) are the distance to the source
$\chi = \sqrt{-K}(\eta_0-\eta)$ and the reduced wavenumber
$\nnu=q/\sqrt{-K}$ (see Appendix \ref{sec:radial} for explicit forms).

The interpretation of these equations is also readily apparent from their form
and construction.
The decomposition of $\Spin{G}{s}{j}{m}$ into radial and spherical parts
encapsulates the summation of spin and orbital angular momentum as well as
the geodesic deviation factors described in \S\ref{sec:normal}.
The difference between the integral solution and the differential form
is that in the former case the coupling is performed in one step from the
source at time $\eta$ and distance $\chi(\eta)$ to the present, while in the
latter the power is steadily transferred to higher $\ell$ as the time advances.

Take the flat space case. 
The intrinsic local angular momentum at the point $(\chi,\hat{n})$ is
$\Spin{Y}{s}{j}{m}$ but must be added to the orbital angular momentum
from the plane wave which can be expanded in terms of $j_\ell Y_\ell^0$.
The result is a sum of $|\ell-j|$ to $\ell+j$ angular momentum states with
weights given by Clebsch-Gordan coefficients.  Alternately a state of definite
angular momentum involves a sum over the same range in the spherical Bessel
function.
These linear combinations of Bessel functions are exactly the radial functions
in Eq.~(\ref{eqn:los}) for the flat limit \cite{TAMM}.

For an open geometry, the same analysis follows save that the spherical
Bessel function must be replaced by a hyperspherical Bessel function
(also called ultra-spherical Bessel functions) in the manner described
in Appendix \ref{sec:radial}.  The qualitative aspect of this modification
is clear from considering the angular diameter distance arguments of
\S\ref{sec:normal}.
The peak in the Bessel function picks out the angle which a scale
$k^{-1} \approx \sqrt{-K}\nnu^{-1}$ subtends at distance
$d \approx \chi/\sqrt{-K}$.
A spherical Bessel function peaks when its argument $kd \approx \ell$
or $\lambda/d \approx \theta$ in the small angle approximation.
The hyperspherical Bessel function peaks at
$k{\cal D}=\nu\sinh\chi \approx \ell$ for $\nu \gg 1$ or
$\lambda/{\cal D} \approx \theta$ in the small angle approximation.
The main effect of spatial curvature is simply to shift features in
$\ell$-space with the angular diameter distance, i.e.~to higher $\ell$
or smaller angles in open universes.
Similar arguments hold for closed geometries \cite{WhiSco}.
By virtue of this fact the division of polarization into $E$ and $B$-modes
remains the same as that in flat space.  More specifically, for a single
mode the ratio in power is given by
\begin{equation}
{\sum_\ell [\ell \beta_\ell^{(m)} ]^2
\over
 \sum_\ell [\ell \epsilon_\ell^{(m)} ]^2 }
= \cases{
        0, & $m= 0, $ \cr
        6, & $m= \pm 1,$ \cr
        8/13, & $m= \pm 2,$ \cr}
\label{eqn:ebratio}
\end{equation}
at fixed source distance $\chi$ with $\nu \sin_K \chi \gg 1$.  

The integral solutions (\ref{eqn:los}) are the basis of the ``line of sight''
method \cite{LOS,SelZal} for rapid numerical calculation of CMB spectra,
which has been employed in CMBFAST.
The numerical implementation of equations (\ref{eqn:los}) requires an efficient
way of calculating the radial functions ($\phi_\ell,\epsilon_\ell,\beta_\ell$).
This is best done acting the derivatives of the hyperspherical Bessel function
in the radial equations~(\ref{eqn:phiradial})-(\ref{eqn:betaradial}) and
(\ref{eqn:phiradialaux}) on the sources through integration by parts.   
The remaining integrals can be efficiently calculated with the techniques of
\cite{ZalSelBer} for generating hyperspherical Bessel functions.
The tensor CMBFAST code has now been modified to use the formalism described
in this paper and the results have been cross-checked against solutions of
the Boltzmann hierarchy equations (\ref{eqn:boltz})-(\ref{eqn:boltzpol}) with
very good agreement.

\subsection{Power Spectra}

The final step in calculating the anisotropy spectra is to integrate over
the $k$-modes.  The power spectra of temperature and polarization anisotropies
today are defined as,
e.g.~$C_\ell^{\Theta\Theta}\equiv\left\langle | a_{\ell m} |^2\right\rangle$
for $\Theta = \sum a_{\ell m} Y_\ell^m$ with the average being over the
($2\ell+1$) $m$-values.  In terms of the moments of the previous section,
\begin{equation}
(2\ell+1)^2 C_\ell^{X\widetilde X} = {2 \over \pi}
        \int {dq \over q} \sum_{m=-2}^2
        q^3 \ X_\ell^{(m)*} \widetilde X_\ell^{(m)}\, ,
\label{eqn:cldef}
\end{equation}
where $X$ takes on the values $\Theta$, $E$ and $B$ for the temperature,
electric polarization and magnetic polarization evaluated at the present.
For a closed geometry, the integral is replaced by a sum over
$q/|K|=3,4,5\ldots$
Note that there is no cross correlation $C_\ell^{\Theta B}$ or $C_\ell^{E B}$
due to parity.  

We caution the reader that power spectra for the metric fluctuation sources
$P_h(q) = \left\langle h^*(q) h(q) \right\rangle$ must be defined in a similar
fashion for consistency and choices between various authors differ by factors
related to the curvature (see \cite{OpenInflation} for further discussion). 
To clarify this point, the initial power spectra of the metric fluctuations
for a scale-invariant spectrum of scalar modes and minimal inflationary
gravity wave modes \cite{OpenTen} are
\begin{eqnarray}
 P_{\Phi}(q) &\propto& {\displaystyle{1 \over \nu(\nu^2+1)}}\,, \nonumber\\
 P_{H}(q) &\propto& {\displaystyle{(\nu^2+4)\over \nu^3(\nu^2+1)}} \tanh(\pi \nu/2)\,,
\label{eqn:initialpower}
\end{eqnarray}
where the normalization of the power spectrum comes from the underlying
theory for the generation of the perturbations.  This proportionality constant
is related to the amplitude of the matter power spectrum on large scales or
the energy density in long-wavelength gravitational waves \cite{OpenInflation}.
The vector perturbations have only decaying modes and so are only present in
seeded models.
The other initial conditions follow from detailed balance of the evolution
equations and gauge transformations (see Appendix \ref{sec:einstein}).

Our conventions for the moments also differ from those in
\cite{SelZal,KamKosSte}.
They are related to those of \cite{SelZal} by\footnote{Footnote 3 of
\cite{TAMM} incorrectly gives the relation between $\Theta$ and $\Delta_T$.}
\begin{eqnarray}
(2\ell+1)\Delta_{T\ell}^{(S)} &=& \Theta_\ell^{(0)}/(2\pi)^{3/2}\,, \nonumber\\
(2\ell+1)\Delta_{T\ell}^{(T)} &=& \sqrt{2}\Theta_\ell^{(2)}/(2\pi)^{3/2}\,,
\end{eqnarray}
where the factor of $\sqrt{2}$ in the latter comes from the quadrature sum
over equal $m=2$ and $-2$ contributions.
Similar relations for $\Delta_{(E,B)\ell}^{(S,T)}$ occur but with an extra
minus sign so that $C_{C,\ell}=-C_\ell^{\Theta E}$ with the other power
spectra unchanged.
The output of CMBFAST {\it continues\/} to be $C_{C,\ell}$ with the sign
convention of \cite{SelZal}.
In the notation of \cite{KamKosSte}, the temperature power spectra agree
but for polarization $C^{EE,BB}_\ell=C^{G,C}_\ell/2$ and
$C^{\Theta E}_\ell=-C^{TG}_\ell/\sqrt{2}$.

\begin{figure*}
\begin{center}
\leavevmode
\epsfxsize=6in \epsfbox{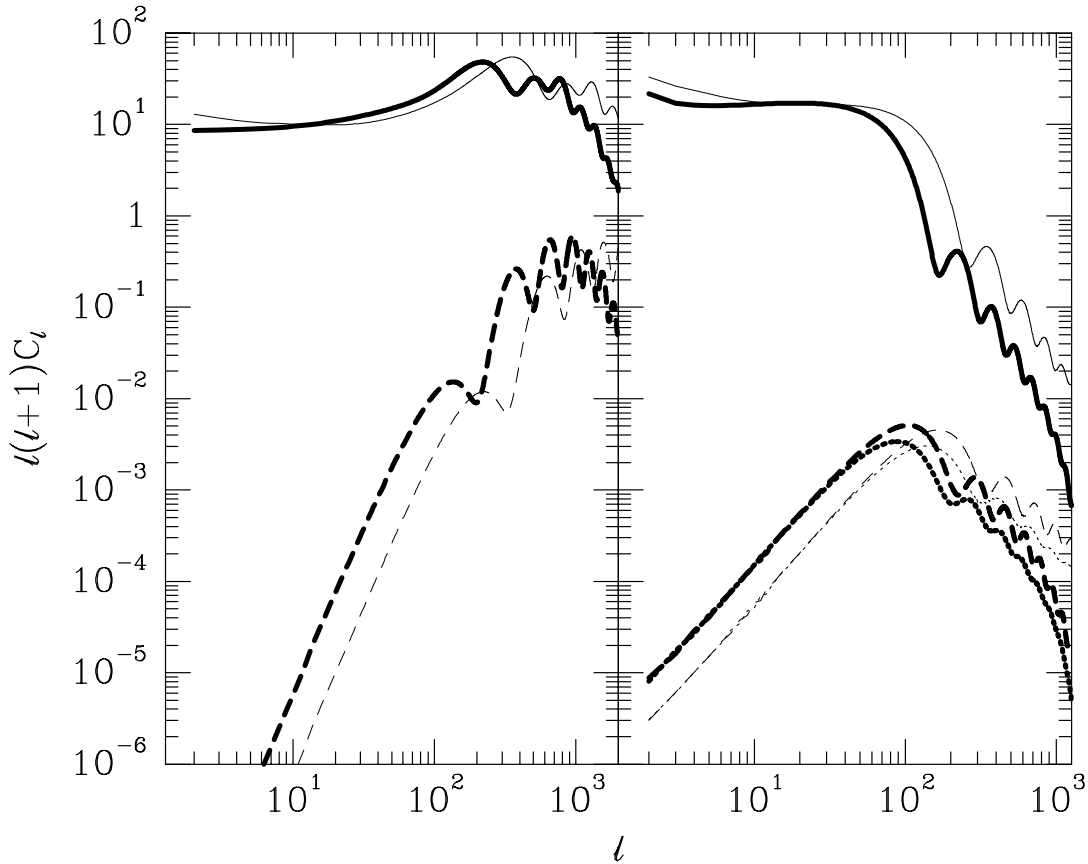}
\end{center}
\caption{The scalar (left) and tensor (right) angular power spectra
for anisotropies in a critical density model (thick lines) and an open
model (thin lines) with $\Omega_0=0.4$.
Solid lines are $C_\ell^{\Theta\Theta}$, dashed $C_\ell^{EE}$ and
dotted $C_\ell^{BB}$.}
\label{fig:temp}
\end{figure*}

\section{Results}
\label{sec:discussion}

We now employ the formalism developed here to calculate the scalar and tensor 
temperature and polarization power spectra for two CDM models one with
critical density and one with $\Omega_0=1-\Omega_K=0.4$ with initial
conditions given by Eq.~(\ref{eqn:initialpower}).
In general, there are two classes of effects: the geometrical and dynamical
aspects of curvature.  

On intermediate to small scales (large $\ell$), only geometrical aspects of
curvature affect the spectra.
Changes in the angular diameter distance to last scattering move features
in the low-$\Omega_0$ models to smaller angular scales (higher $\ell$) as
discussed in \S\ref{sec:boltzmann}.
Since the low-$\ell$ tail of the $E$-mode polarization is growing rapidly
with $\ell$, shifting the features to higher $\ell$ results in smaller
large-angle polarization in an open model for both scalar and tensor
anisotropies.  The suppression is larger in the case of scalars than tensors
since the low-$\ell$ slope is steeper \cite{TAMM}.  

\begin{figure*}
\begin{center}
\leavevmode
\epsfxsize=6in \epsfbox{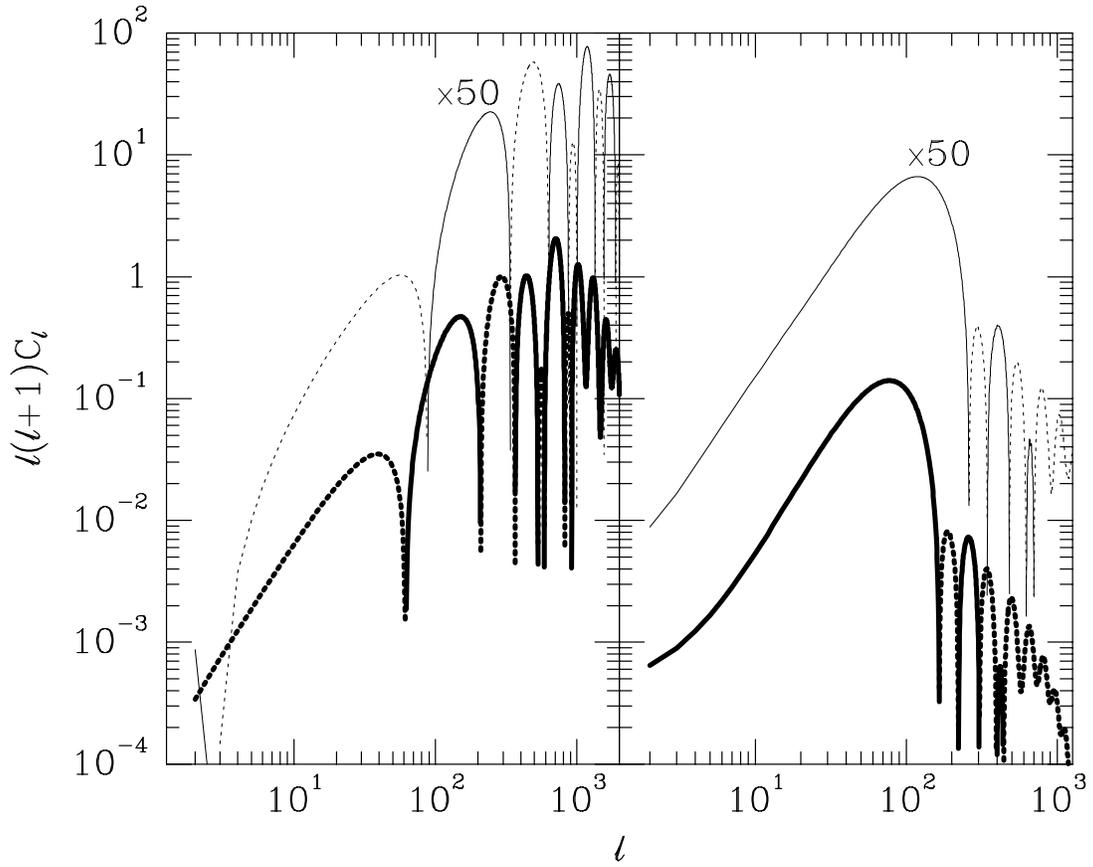}
\end{center}
\caption{The scalar (left) and tensor (right) temperature-polarization 
cross correlation $C_\ell^{\Theta E}$ with the same parameters and notation
as Fig.~\protect\ref{fig:temp} (thick: flat; thin open). Dotted lines
represent negative correlation.}
\label{fig:cross}
\end{figure*}
The presence of curvature also affects the late-time dynamics and initial
power spectra.  As is well known, the scalar temperature power spectrum
exhibits an enhancement of power at low multipoles due to the integrated
Sachs-Wolfe (ISW) effect during curvature domination.
This does not affect the polarization, assuming no reionization, as it is
generated at last scattering.  However it {\it does\/} affect the
temperature-polarization cross correlation (see Fig.~\ref{fig:cross}).
In an open universe, the largest scales (lowest $\ell$) pick up unequal-time
correlations with the ISW contributions which are of opposite sign to the
ordinary Sachs-Wolfe contribution.
This reverses the sign of the correlation and formally violates the
predictions of \cite{CriCouTur}.
In practice this effect is unobservable due to the smallness of signal.
Even minimal amounts of reionization will destroy this effect.

Open universe modifications to the initial power spectrum are potentially
observable in the large angle CMB spectrum.
Unfortunately subtle differences in the temperature power spectrum can be
lost in cosmic variance.  
While polarization provides extra information, in the absence of late
reionization the large-angle polarization is largely a projection of small-scale 
fluctuations.
Nonetheless in our universe (where reionization occured before redshift
$z\approx 5$) the large-angle polarization is sensitive to the primordial
power spectrum at the curvature scale.
Thus if the fluctuations which gave rise to the large-scale structure and
CMB anisotropy in our universe were generated by an open inflationary scenario
based on bubble nucleation, a study of the large-angle polarization can in
principle teach us about the initial nucleation event \cite{OpenInflation}.

\vskip 0.5truecm
In summary, we have completed the formalism for calculating and interpreting
temperature and polarization anisotropies in linear theory from arbitrary
metric fluctuations in an FRW universe.
The results presented here are new for non-flat vector and tensor
(polarization) perturbations and we have calculated the scalar and tensor
temperature and polarization contributions for open inflationary spectra. 
The open tensor perturbation equations have been added to CMBFAST which is now
publically available.

\vskip 0.5truecm
\noindent
{\it Acknowledgments:} We thank the Aspen Center for Physics
where a portion of this work was completed.  W.H was supported
by the W.M. Keck Foundation and NSF PHY-9513835; 
M.Z. by NASA Grant NAG5-2816.

\appendix

\section{Einstein Equations} 
\label{sec:einstein}

In this Appendix, we complete the Boltzmann equations of \S\ref{sec:boltzmann}
by giving the Einstein equations for the metric and the matter.
We begin with the background evolution and then proceed to the fluctuations.
It is occasionally convenient to shift between different representations or
gauges and thus we first discuss the transformations that link them.
We then derive and present the Einstein equations for scalar, vector and
tensor perturbations in a universe with constant comoving curvature in the
synchronous and Newtonian gauges (see also \cite{HSW}).

\subsection{Background Evolution}

The Einstein equations
$G_{\mu\nu} = 8\pi G T_{\mu\nu}$ express the metric evolution
in terms of the matter sources.  The background evolution equations
are
\begin{eqnarray}
{\dot \rho_f \over \rho_f} + 3(1+w_f) {\dot a \over a} = 0 \,,\nonumber\\
{\ddot \phi} + 2{\dot a \over a} \dot\phi +a^2 \scpot_{,\phi} = 0\,,
\end{eqnarray}
for the fluid and scalar field components respectively and
\begin{eqnarray}
\left( {\dot a \over a} \right)^2 + K = {8\pi G \over 3} a^2 (
\rho_f + \rho_\phi + \rho_v) \,,
\end{eqnarray}
where $w_f=p_f/\rho_f$ and 
$\rho_\phi(\phi)$ was given in Eq.~(\ref{eqn:scalarfieldbac})
and $\rho_v =3 H_0^2 \Omega_\Lambda / 8\pi G$ is the vacuum energy.

\subsection{Gauge Transformations} 
\label{sec:gauge}

To represent the perturbations we must make a gauge choice.
A gauge transformation is a change in the correspondence between the
perturbation and the background  represented by the coordinate shifts
\begin{equation}
\begin{array}{rcl}
\tilde \eta &=& \eta + TQ^{(m)}, \nonumber\\
\tilde x_i  &=& x_i + L Q_i^{(m)}.
\end{array}
\label{eqn:shift}
\end{equation}
$T$ corresponds to a choice in time slicing and $L$ a choice of
spatial coordinates.  Since scalar and vector quantities 
cannot be formed from tensor modes ($m=\pm 2$),
no gauge freedom remains there.  Under the condition that
metric distances be invariant,
they transform the metric as \cite{KodSas}
\begin{eqnarray}
\tilde A^{(m)} &=& A^{(m)} - \dot T - {\dot a \over a} T\,, \nonumber\\
\tilde B^{(m)} &=& B^{(m)} + \dot L + kT\,, \nonumber\\
\tilde H_L^{(m)} &=& H_L^{(m)} - {k \over 3}L - {\dot a \over a} T\,, 
	\nonumber\\
\tilde H_T^{(m)} &=&  H_T^{(m)} + kL\,.
\label{eqn:metrictrans}
\end{eqnarray}

The stress-energy perturbations in different gauges are similarly
related by the gauge transformations
\begin{eqnarray}
\tilde \delta_f &=& \delta_f + 3(1+w_f){\dot a \over a} T\,, \nonumber\\
\tilde{\delta p_f}  &=& \delta  p_f + 3c_{f}^2 \rho_f
(1+w_f){\dot a \over a} T\,, \nonumber\\
\tilde v_f^{(m)} &=& v_f^{(m)} + \dot L\,, \nonumber\\
\tilde \pi_f^{(m)} &=& \pi_f^{(m)}\,.
\label{eqn:fluidtrans}
\end{eqnarray}
Note that the anisotropic stress is gauge-invariant.
Seed perturbations are also gauge-invariant to lowest order, whereas
a scalar field transforms as
\begin{equation}
\tilde{\delta\phi} = \delta \phi - {\dot \phi} T \,.
\end{equation}
The relation between the synchronous and Newtonian gauge equations follow 
from these relations.

\subsection{Scalar Einstein Equations} \label{sec:scalar}

With the form of the scalar metric and stress energy tensor given
in Eqs.~(\ref{eqn:metrictrans}) and (\ref{eqn:stress}), the ``Poisson''
equations become in the Newtonian gauge
\begin{equation}
\begin{array}{rcl}
(k^2-3K) \Phi &=& 4\pi G a^2 \left[ (\rho_f \delta_f + \rho_s) + 3
\displaystyle{\dot a \over a}[ (\rho_f + p_f)v_f^\se + v_s^\se]
/k \right] , \\
k^2 (\Psi + \Phi) &=& -8\pi G a^2 \left( p_f \pi_f^\se + \pi_s^\se \right),
\vertsp
\end{array}
\label{eqn:Poisson}
\end{equation}
and in the synchronous gauge
\begin{eqnarray}
(k^2 - 3K)(h_L + {1 \over 3} h_T) + 3{\dot a \over a} \dot h_L
& = & 4\pi G a^2[\rho_f \delta_f + \rho_s ] \nonumber\,, \\
\dot h_L + {1 \over 3}(1-3K/k^2) \dot h_T &=& 
	-4\pi G a^2 [(\rho_f + p_f)v_f^\se + v_s^\se]/k \nonumber\,, \\
\ddot h_L + {\dot a \over a} \dot h_L &=&
	-4\pi G a^2 [{1 \over 3}\rho_f \delta_f + \delta p_f
	+ {1 \over 3} \rho_s + p_s] \nonumber\,, \\
\ddot h_T + {\dot a \over a}\dot h_T - k^2(h_L + {1 \over 3}h_T) &=&
	-8\pi Ga^2 [p_f \pi_f^\se + \pi_s^\se]\,.
\end{eqnarray}
Two out of four of the synchronous gauge equations are redundant.

The corresponding evolution of the matter is given by covariant conservation
of the stress energy tensor $T_{\mu\nu}$: 
\begin{eqnarray}
\dot \delta_f
& = & -(1+w_f)kv_f^\se - 3{\dot a \over a} \delta w_f + S_\delta 
        \, , \nonumber\\
\left[ (1 + w_f) v_f^\se \right]\dot{\vphantom{\Big[}}
& = & -(1+w_f)  {\dot a  \over a} (1- 3w_f) v_f^\se
        + w_f k\left[\delta p_f/p_f-{2\over 3}(1-3K/k^2)\pi_f^\se \right] + S^{(0)}_v \, ,
\label{eqn:sfluideqnN}
\end{eqnarray}
for the fluid part (see e.g. \cite{Pee}).  The gravitational sources are
\begin{equation}
\begin{array}{lll}
S_\delta  = -3(1+w_f)\dot \Phi\,, \quad& S_v^{(0)} =
   (1+w_f) k\Psi\,,\qquad & (\rm Newtonian),\vphantom{\Big[} \\
S_\delta = -3(1+w_f)\dot h_L\,,   \quad& S_v^{(0)} =
   0\,,\qquad & (\rm synchronous) \vphantom{\Big[}.
\end{array}
\end{equation}
These equations remain true for each fluid individually in the absence
of momentum exchange, e.g.~for the cold dark matter.
The baryons have an additional term to the Euler equation due to momentum
exchange from Compton scattering with the photons.
For a given velocity perturbation the momentum density ratio between the two
fluids is
\begin{equation}
R \equiv {\rho_B+p_B \over \rho_\gamma + p_\gamma} \approx
{3\rho_B \over 4\rho_\gamma} \, .
\label{eqn:Rdef}
\end{equation}
A comparison with photon Euler equation~(\ref{eqn:boltz}; $\ell=1$) gives
the source modification for the baryon Euler equation
\begin{equation}
S_v^{(0)} \rightarrow S_v^{(0)} + {\dot\tau \over R}
        (\Theta_1^\se - v_B^\se)\, .
\label{eqn:baryonscat}
\end{equation}
For a seed source, the conservation equations become
\begin{eqnarray}
\dot \rho_s &=&
        -3{\dot a \over a} (\rho_s + p_s) - kv_s^\se , \nonumber\\
\dot v_s^\se &=&
        -4{\dot a \over a} v_s^\se + k\left[ p_s -{2 \over 3}
	(1-3K/k^2)\pi_s^\se \right]\, ,
\end{eqnarray}
independent of gauge since the metric fluctuations produce higher order terms.

Finally for a scalar field, $\varphi=\phi+\delta\phi$, the conservation
equations become
\begin{equation}
\ddot{\delta\phi} + 2{\dot a \over a} \dot{\delta\phi} 
+ (k^2 + a^2 \scpot_{,\phi\phi})\delta\phi = S_\phi \,,
\end{equation}
where
\begin{equation}
S_\phi = \cases{ (\dot\Psi-3\dot\Phi)\dot\phi -
   2 a^2 \scpot_{,\phi}\Psi \,,& (Newtonian), 
\vphantom{\Big[} \cr
   -3\dot h_L \dot\phi\,, & (synchronous),
\vphantom{\Big[}
\cr }
\end{equation}
are the gravitational sources. 

\subsection{Vector Einstein Equations} \label{sec:vector}

The vector metric source evolution is similarly constructed from
a ``Poisson'' equation: in the generalized Newtonian gauge
\begin{equation}
\dot V + 2 {\dot a \over a} V =
        -8\pi G a^2 (p_f \pi^\ve_f + \pi_s^\ve)/k \, ,
\end{equation}
and for the synchronous gauge,
\begin{equation}
\ddot h_V + 2 {\dot a \over a} \dot h_V =
        -8\pi G a^2 (p_f \pi^\ve_f + \pi_s^\ve)/k^2 \, .
\end{equation}
Likewise momentum conservation implies the Euler equation
\begin{equation}
 \dot v_f^\ve =  - (1 - 3c_f^2 )
{\dot a \over a} v^\ve_f  - {1 \over 2 } k
{w_f \over 1+w_f} (1 - 2K/k^2)\pi^\ve_f + S_v^{(1)},
\label{eqn:vectoreulersyn}
\end{equation}
where recall $c_f^2=\dot{p}_f/\dot{\rho}_f$ is the sound speed and the
gravitational sources are
\begin{equation}
S_v^{(1)} = \cases {
  \dot V + (1-3c_f^2) {\displaystyle{\dot a \over a}} V\,,
    & (Newtonian), \cr
  0\,, & (synchronous). \cr}
\end{equation}
The seed Euler equation is given by
\begin{equation}
\dot v_s^\ve = - 4{\dot a \over a} v_s^\ve  -
        {1 \over 2}k(1-2K/k^2)\pi_s^\ve.
\end{equation}
Again, the first of these equations remains true for each fluid
individually save for momentum exchange terms.  The baryon Euler equation has
an additional term in the source of the same form as 
Eq.~(\ref{eqn:baryonscat}) with $m=0 \rightarrow m=1$.

\subsection{Tensor Einstein Equations} \label{sec:tensor}

The Einstein equations tell us that the tensor metric source is governed by
\begin{equation}
\ddot H + 2{\dot a \over a} \dot H +
(k^2 + 2K) H = 8\pi G a^2 [
p_f^{\vphantom{\te}} \pi^{\te}_f + \pi^{\te}_s]\, ,
\end{equation}
for all gauges.

\section{Radial Functions}
\label{sec:radial}

It is often useful to represent the eigenmodes 
in a spherical coordinate system $(\chi,\theta,\phi)$, where $\chi$
is the radial coordinate scaled to the curvature radius.  Here we
explicitly write down the forms and properties of the radial
modes in an open geometry and describe the modifications 
necessary to treat closed geometries.

By separation of variables in the Laplacian, we can write 
\begin{equation}
\Spin{G}{s}{j}{m}
        =   \sum_\ell (-i)^\ell \sqrt{4\pi(2\ell+1)} \,
          \Spin{\alpha}{s}{\ell}{(j m)}(\chi,\nnu) 
	   \, \Spin{Y}{s}{\ell}{m}({\hat{n}})  \, ,
\label{eqn:radialdecomposition}
\end{equation}
and the goal is to find explicit expressions for 
$\Spin{\alpha}{s}{\ell}{(j m)}$.  Here the $\ell$-weights are set to reproduce
the flat-space conventions of spherical Bessel functions (see also \cite{TAMM}).
We proceed by analyzing the lowest $j = {\rm min}(|s|,|m|)$ harmonic
\begin{eqnarray}
\Gm{0}{j}{m} &=& n^{i_1}\ldots n^{i_{|m|}}  
	Q_{i_1\ldots i_{|m|}}^{(m)} \,,   
	\nonumber \\
\Gm{\pm 2}{2}{m} & \propto & (\hat{m}_1 \pm i \hat{m}_2)^{i_1}
	(\hat{m}_1 \pm i \hat{m}_2)^{i_2} 
Q_{i_1 i_2}^{(m)} \,, 
\label{eqn:Gl0}
\end{eqnarray}
where $\hat{m}_1$ and $\hat{m}_2$ form a right-handed orthonormal 
basis with $\hat{n}$.
We can now determine $\Spin{\alpha}{s}{\ell}{(j m)}$ from
the radial representation of ${\bf Q}^{(m)}$ \cite{Tom}
\begin{eqnarray}
\phi_{\ell}^{(00)}(\chi,\nnu) 
	& = & \Phi_\ell^\nnu(\chi) \, , \nonumber\\
\phi_{\ell}^{(11)}(\chi,\nnu)
&=&\sqrt{\ell(\ell+1)\over2(\nnu^2+1)} {\rm csch}\chi\,\Phi_\ell^\nnu(\chi)\,,
	\nonumber\\
\phi_{\ell}^{(22)}(\chi,\nnu)
& = & \sqrt{{3 \over 8}{(\ell+2)(\ell^2-1)\ell \over (\nnu^2+4)(\nnu^2+1)}}
	{\rm csch}^2\chi\, \Phi_\ell^\nnu(\chi) \,,
\label{eqn:phiradial}
\end{eqnarray}
for $\Spin{\alpha}{0}{\ell}{(m m)}=\phi_\ell^{(m m)}$;
similarly for 
$\Spin{\alpha}{\pm 2}{\ell}{(2 m)}=\epsilon_\ell^{(m)}\pm i\beta_\ell^{(m)}$,
\begin{eqnarray}
\epsilon_{\ell}^{(0)}(\chi,\nnu)
& = & \sqrt{{3 \over 8}{(\ell+2) (\ell^2-1)\ell \over (\nnu^2+4)(\nnu^2+1)}}
	{\rm csch}^2\chi \Phi_\ell^\nnu(\chi) \,,\nonumber\\
\epsilon_{\ell}^{(1)}(\chi,\nnu)
& = & {1 \over 2} \sqrt{(\ell-1)(\ell+2) \over (\nnu^2 +4)(\nnu^2+1)}
{\rm csch}\chi \left[ {\rm coth}\chi \Phi_\ell^\nnu(\chi) 
+ \Phi_\ell^\nnu{}'(\chi) \right]
\,,\nonumber\\
\epsilon_{\ell}^{(2)}(\chi,\nnu)
	& = & {1 \over 4} \sqrt{1 \over (\nnu^2+4)(\nnu^2 + 1)}
\left[ \Phi_\ell^\nnu{}''(\chi) + 4 {\rm coth}\chi \Phi_\ell^\nnu{}' (\chi)
- \left( \nnu^2 -1 - 2 {\rm coth}^2 \chi \right) \Phi_\ell^\nnu(\chi)
 \right]\,,
\label{eqn:epsilonradial}
\end{eqnarray}
and 
\begin{eqnarray}
\beta_{\ell}^{(0)}(\chi,\nnu) &=& 0 \,, \nonumber \\
\beta_{\ell}^{(1)}(\chi,\nnu) &=& 
{1 \over 2} \sqrt{(\ell-1)(\ell+2)\nnu^2 \over (\nnu^2 +4)(\nnu^2+1)}
	{\rm csch}\chi \Phi_\ell^\nnu(\chi) \,, \nonumber \\
\beta_{\ell}^{(2)}(\chi,\nnu) &=&
{1 \over 2}\sqrt{\nnu^2 \over (\nnu^2+4)(\nnu^2 + 1)}
\left[ \Phi_\ell^\nnu {}'(\chi)+2{\rm coth}\chi\Phi_\ell^\nnu(\chi) \right]\,,
\label{eqn:betaradial}
\end{eqnarray}
for $m > 0$. For $m<0$, $\beta_\ell^{(-m)} = -\beta_\ell^{(m)}$ while the other
two functions remain the same.
Here $\Phi_\ell^\nnu(\chi)$ is the hyperspherical Bessel function
whose properties are discussed extensively by \cite{AbbSch}. 

The overall normalization of the modes here has been altered from those of
\cite{AbbSch,Tom} in the case of vector and tensor temperature modes such that
\begin{equation}
\Spin{\alpha}{s}{\ell}{(j m)}(0,\nnu) = {1 \over 2\ell+1}\delta_{\ell,j} \, ,
\end{equation}
where the difference lies in the lack of curvature dependence in the relation.
Our choice simplifies the equations since it preserves the flat space form of
the equations locally around the origin.
It also {\it defines\/} the normalization of the polarization modes with
respect to $Q_{ij}^{(m)}$ through Eq.~(\ref{eqn:Gl0}).

The properties of the hyperspherical Bessel functions imply useful properties
for the radial functions.
For our purposes, the important relations they obey are:
\begin{eqnarray}
{d \over d\chi}\Phi_\ell^\nnu & = &
{1 \over 2\ell+1} \left[ \ell \sqrt{\nnu^2 + \ell^2} \Phi_{\ell-1}^\nnu
	- (\ell+1)\sqrt{\nnu^2 + (\ell+1)^2} \Phi_{\ell+1}^\nnu \right]
	\nonumber\,,\\
{\rm coth}\chi \Phi_\ell^\nnu & = &
{1 \over 2\ell+1} \left [ \sqrt{\nnu^2 + \ell^2} \Phi_{\ell-1}^\nnu
	+ \sqrt{\nnu^2 + (\ell+1)^2} \Phi_{\ell+1}^\nnu \right]\,,
\end{eqnarray}
which define the series in terms of its first member
\begin{equation}
\Phi_0^\nnu = {\sin{\nnu\chi} \over \nnu\sinh{\chi}}.
\end{equation}
Notice that $\lim_{K \rightarrow 0} \Phi_\ell^\nnu(\chi) = j_\ell(kr)$.

From the recursion relations of $\Phi_\ell^\nnu$, one establishes
the corresponding relations for the radial function
\begin{equation}
{d \over d\chi} [\Spin{\alpha}{s}{\ell}{(j m)}]
	= {\nnu \over 2\ell +1} \left\{
	  {\Spin{\kappa}{s}{\ell}{m}}  
	  \left[\Spin{\alpha}{s}{\ell-1}{(j m)}\right]
	- {\Spin{\kappa}{s}{\ell+1}{m}} 
	  \left[\Spin{\alpha}{s}{\ell+1}{(j m)}\right]
	\right\}
	- {\nnu m s \over \ell(\ell+1)} 
	  \left[\Spin{\alpha}{s}{\ell}{(j m)}\right]\, ,
\label{eqn:radialrecursionA}
\end{equation} 
for the lowest $j$, where recall 
\begin{equation}
\Spin{\kappa}{s}{\ell}{m}
 =\sqrt{\left[ {(\ell^2-m^2)(\ell^2-s^2)\over \ell^2}\right]
\left[ 1+{\ell^2\over\nnu^2}\right] }\, . 
\end{equation}

The construction of the higher $\Spin{G}{s}{\ell}{m}$ via
the recursion relation of Eq.~(\ref{eqn:recursion}) also returns
the higher radial harmonics.  A few useful ones are
\begin{eqnarray}
\phi_{\ell}^{(10)}(\chi,\nnu)
& = & \sqrt{1 \over \nnu^2 + 1} \Phi_\ell^\nnu{}'(\chi) \, , \nonumber\\
\phi_{\ell}^{(20)}(\chi,\nnu)
& = & {1 \over 2} \sqrt{1 \over (\nnu^2+4)(\nnu^2+1)}
\left[ 3 \Phi_\ell^\nnu{}''(\chi) + (\nnu^2+1)\Phi_\ell^\nnu (\chi)
	\right] \, , \nonumber\\
\phi_{\ell}^{(21)}(\chi,\nnu)
	& = &\sqrt{ {3 \over 2} {\ell(\ell+1) \over (\nnu^2+4)(\nnu^2+1) }} 
	   \left[ {\rm csch}\chi \Phi_\ell^\nnu(\chi) \right]' \,. 
\label{eqn:phiradialaux}
\end{eqnarray}
Furthermore, the recursion relation obeyed by the higher radial harmonics
is the same as Eq.~(\ref{eqn:radialrecursionA}), by virtue of
Eq.~(\ref{eqn:recursionA}) and explicit substitution of the radial form
Eq.~(\ref{eqn:radialdecompositionC}).
This $j$-independence of the recursion relation implies that
$\phi_\ell^{(j m)}$ is a solution to the temperature hierarchy
Eq.~(\ref{eqn:boltz}) for any $j$ and aids in the construction of
the integral solutions in \S\ref{sec:integral}.

Finally, the radial functions for a closed geometry follow by replacing
all $\nnu^2 + n$, where $n$ is integer, with $\nnu^2 - n$ and trigonometric 
functions with hyperbolic trigonometric functions (see \cite{AbbSch,Tom} for
details).  

\section{Derivation of the Normal Modes} \label{sec:derivation}

We would like to describe the spatial and angular dependence of the normal
modes $\Spin{G}{s}{\ell}{m}(\vec{x},\hat{n})$ in a coordinate-free way by
constructing them out of covariant derivatives of ${\bf Q}^{(m)}$ contracted
with some orthonormal basis $(\hat{n},\hat{m}_1,\hat{m}_2)$.
The lowest $j={\rm max}(|m|,|s|)$ modes can be written as
\cite{OpenTen,ZalSelBer},  
\begin{eqnarray}
\Gm{0}{j}{m} &=& n^{i_1}\ldots n^{i_{|m|}}  
	Q_{i_1\ldots i_{|m|}}^{(m)} \,,   
	\nonumber \\
\Gm{\pm 2}{2}{m} & \propto & (\hat{m}_1 \pm i \hat{m}_2)^{i_1}
	(\hat{m}_1 \pm i \hat{m}_2)^{i_2} 
Q_{i_1 i_2}^{(m)} \,, 
\label{eqn:Gl0prime}
\end{eqnarray}
and satisfy (Appendix \ref{sec:radial}),
\begin{equation}
\Gm{s}{\ell}{m}(\vec{x},\hat{n})  =
        (-i)^\ell \sqrt{ {4\pi \over 2\ell+1}}
        [\Spy{s}{\ell}{m}(\hat{n})] \exp[i\delta(\vec{x},\vec{k})]\,,
\label{eqn:generalizedGA}
\end{equation}
with $\ell=j$.  We demand that the higher $\ell$-modes also do so,
to maintain the division of spin and orbital angular momentum defined in
flat space \cite{TAMM}. 

We begin the construction by choosing some arbitrary point $\vec{x}_0$,
and using a spherical coordinate system around it, 
$\vec{x}-\vec{x}_0=\sqrt{-K}\, \chi(-\hat{n})$.
Now $\hat{n}$ defines both the intrinsic angular coordinate system and
the angular coordinates for the spatial location $\vec{x}(\chi,\hat{n})$.
This reduction in the dimension of the space is sufficient since the
end goal is to derive how the intrinsic and orbital angular dependence
in the same direction $\hat{n}$ adds.  In physical terms, only those
photons directed toward the observer can contribute to the local
angular dependence there.
First expand the lowest mode  in spin-spherical harmonics
\begin{equation}
\Spin{G}{s}{j}{m}(\chi,\hat{n};\nu) = 
	\sum_\ell (-i)^\ell \sqrt{4\pi(2\ell+1)}\,
	\Spin{\alpha}{s}{\ell}{(j m)}(\chi,\nnu) \,
	\Spin{Y}{s}{\ell}{m}({\hat{n}})\, ,
\label{eqn:radialdecompositionC}
\end{equation}
where recall that the dimensionless wavenumber is $\nnu =
q/\sqrt{-K}$.
We obtain the explicit expressions for $\Spin{\alpha}{s}{\ell}{(j m)}$
and their recursion relations
in Appendix \ref{sec:radial} by simple comparison between equations
(\ref{eqn:Gl0prime}) 
and (\ref{eqn:radialdecompositionC}). At the origin they satisfy
\begin{equation}
\Spin{\alpha}{s}{\ell}{(j m)}(0,\nnu) = {1 \over 2\ell+1}\delta_{\ell,j} \, ,
\label{eqn:radialorigin}
\end{equation}
which both fixes the normalization of the modes and manifestly obeys
Eq.~(\ref{eqn:generalizedGA}).
As $\chi \rightarrow 0$, only the local angular dependence remains, as
expressed in the Kronecker delta of Eq.~(\ref{eqn:radialorigin}).
Because the spatial variation of the normal mode $Q^{(m)}$ across a shell
at fixed radius $\chi$ must be added to the local dependence, even a mode
of fixed $j$ has a sum over all $\ell$ in its angular dependence which 
contributes at any other point.

This generation of higher $\ell$-structure as $\chi$ increases suggests that
we can use the radial structure of $\Spin{G}{s}{j}{m}$ to generate the higher
$\ell$-modes.  
{}From the radial recursion relation for $\Spin{\alpha}{s}{\ell}{(jm)}$
Eq.~(\ref{eqn:radialrecursionA}), let us make the ansatz
\begin{equation}
{1 \over \sqrt{-K}} n^i (\Spin{G}{s}{\ell}{m})_{|i} 
        = {\nnu \over 2\ell +1} \left[
          {\Spin{\kappa}{s}{\ell}{m}}  \Spin{G}{s}{\ell-1}{m}
        - {\Spin{\kappa}{s}{\ell+1}{m}} \Spin{G}{s}{\ell+1}{m}
        \right]
        - i{\nnu m s \over \ell(\ell+1)} \Spin{G}{s}{\ell}{m}.
\label{eqn:recursionA}
\end{equation}
That this series generates modes with the desired properties can be shown
by returning to the spherical coordinate system.
By explicit substitution of the radial form for 
$\Spin{G}{s}{j}{m}$ of Eq.~(\ref{eqn:radialdecompositionC})
and by noting that in this coordinate system
\begin{equation}
{1 \over \sqrt{-K}} n^i (\Spin{G}{s}{\ell}{m})_{|i} =
   -{d \over d\chi} (\Spin{G}{s}{\ell}{m})\,,
\label{eqn:angularradial}
\end{equation}
we obtain
\begin{equation}
\Gm{s}{\ell}{m}(0,\hat{n})  =
        (-i)^\ell \sqrt{ {4\pi \over 2\ell+1}}
        [\Spy{s}{\ell}{m}(\hat{n})]  \,,
\label{eqn:originG}
\end{equation}
(up to a phase factor) as desired.
Since we have shown this for an arbitrary point, it is clear that
Eq.~(\ref{eqn:generalizedGA}) holds in general. 
Note that this construction requires 
\begin{equation}
\int {d\Omega \over 4\pi}\, \big| [\Spin{ G}{s}{\ell_1}{m_1}]^*\,
[\Spin{G}{s}{\ell_2}{m_2}] \big|
        = {1 \over 2\ell_1+1}
        \delta_{\ell_1,\ell_2}\, \delta_{m_1 ,m_2} \,,
\label{eqn:gnorm}
\end{equation}
for all $\vec{x}$, as in the flat case of Eq.~(\ref{eqn:flatG}), and defines 
our normalization convention.

\end{document}